# How Retailers at different Stages of E-Commerce Maturity Evaluate Their Entry to E-Commerce Activities?

Rayed Al Ghamdi[1], Osama Abdulaziz Alfarraj[2] & Adel A. Bahaddad[3]

**Abstract**

This paper investigates how retailers at different stages of e-commerce maturity evaluate their entry to e-commerce activities. The study was conducted using qualitative approach interviewing 16 retailers in Saudi Arabia. It comes up with 22 factors that are believed the most influencing factors for retailers in Saudi Arabia. Interestingly, there seem to be differences between retailers in companies at different maturity stages in terms of having different attitudes regarding the issues of using e-commerce. The businesses that have reached a high stage of e-commerce maturity provide practical evidence of positive and optimistic attitudes and practices regarding use of e-commerce, whereas the businesses that have not reached higher levels of maturity provide practical evidence of more negative and pessimistic attitudes and practices. The study, therefore, should contribute to efforts leading to greater e-commerce development in Saudi Arabia and other countries with similar context.

**Keywords**: Retailing, e-commerce, adoption, e-commerce maturity, KSA

## 1. Introduction

A great deal of literature has identified factors that influence businesses on whether to adopt or not to adopt e-commerce (Kendall et al., 2001; Tigre, 2003; Drew, 2003; Andersen, Björn & Dedrick, 2003; Grewal, Iyer & Levy, 2004; Wymer & Regan, 2005; Gibbs, et al., 2006; Hafez, 2006; Sleem, 2006; Shalhoub & AlQasimi, 2006; To & Ngai, 2006; Kraemer, Dedrick, & Melville, 2006; Ho, Kauffman & Liang, 2007; El Said & Galal-Edeen, 2009; Scupola, 2009; Colton, Roth & Bearden, 2010; Nair, 2010).

[1] Faculty of Computing & IT, King Abdulaziz University, Jeddah, Kingdom of Saudi Arabia.
Email: raalghamdi8@kau.edu.sa, Phone: +966 5 5844 1379
[2] Computer Science Department, Community College, King Saud University, Riyadh, Kingdom of Saudi Arabia. Email: oalfarraj@ksu.edu.sa
[3] Faculty of Computing and IT, King Abdulaziz University, Jeddah, Kingdom of Saudi Arabia. Email: dbabahaddad10@kau.edu.sa

These studies dealt with various factors that might influence businesses to adopt and use e-commerce concentrating mainly on organizational and business factors, and, to a lesser extent, customer practices and environmental factors. The concentration on the organizational factors possible relates to the fact that the e-commerce infrastructure environment is already well established in the countries studied. In studies held within Saudi Arabia there is greater emphasis on environmental factors; such as, ICT infrastructure, logistics, and online payment mechanisms (Aladwani, 2003; Albadr, 2003; Sait, Al-Tawil & Hussain, 2004; Al-Solbi & Mayhew, 2005; Alfuraih, 2008; Alwahaishi, Nehari-Talet & Snasel, 2009; Alrawi & Sabry, 2009; Aleid, Rogerson & Fairweather, 2010). Thus, for example, logistic infrastructure issues, such as the lack of mailing addresses and the cost of delivery in Saudi Arabia, have been highlighted as inhibiting factors for businesses to sell online. Little attention has been given to non environmental factors. Moreover, no studies have investigated how the Saudi businesses that already sell online and have reached a high level of e-commerce maturity overcame these issues. A study of companies at different levels of e-maturity could be invaluable in suggesting what could be done to increase the likelihood of other company's increasing their maturity stage.

## 2. Literature Review

2.1. E-commerce and Saudi Arabia

At the most basic level, commerce is the trade of goods for money, and electronic commerce (e-commerce) is commerce enabled by the Internet (Whiteley, 2000). Comprehensively, e-commerce includes pre-sale and post-sale activities across the supply chain over the Internet (Chaffey, 2004). It can be defined as "all electronically mediated information exchanges between an organization and its external stakeholders" (Chaffey, 2004, p. 9). In some definitions, e-commerce is made equivalent to electronic business (e-business). However, in this study, e-commerce is considered as a subset of e-business in line with Davis & Benamati (2003).

E-business can be defined as business processes using "all electronically mediated information exchanges, both within an organization and with external stakeholders" (Chaffey, 2004, p. 10). As the present study investigates what would encourage online retailing in KSA, the definition that defined e-commerce as commerce enabled by Internet including pre-sale and post-sale activities was adopted in this research. This definition facilitates the explanation of this process.

E-commerce encompasses a broad area and is classified into different subtypes or 'models'. The classification encompassing Business to Customer (B2C), Business to Business (B2B), Customer to Customer (C2C), Peer to Peer (P2P), and Mobile Commerce (M-commerce). Davis & Benamati (2003) categorize e-commerce into B2C, B2B, C2C, and Business to Employee (B2E). There are also Business to Government (B2G) and Consumer to Government (C2G) models to consider. The focus of this study falls under B2C.

B2C is categorized into seven models (Laudon & Traver, 2010); portal, online retailer, content provider, transaction broker, market creator, service provider, and community provider. To narrow the research topic further down, this research is involved with online retailing. Online retailing can be defined as an "online version of traditional retail; [which] includes virtual merchant (online retail store only), Bricks-and-Clicks e-retailers (online distribution channel for a company that also has physical store), Catalog Merchants (online version of direct mail catalog), Manufacturers selling directly over the web" (Laudon & Traver, 2010).

Many businesses around the world have introduced e-commerce tools into their businesses to gain a competitive advantage. The adoption of e-commerce systems, especially in the developed continues, is growing quickly (Kamaruzaman, Handrich & Sullivan, 2010). The period 1995-2000 saw a notable proliferation of e-commerce start-ups and online retailing systems in the USA (Dedrick et al., 2006; Dinlersoz & Hernández-Murillo, 2005). Since 2000, the rapid growth of e-commerce activities has been obvious in the developed world. Global e-commerce spending is worth about US$10 trillion at present, compared to US$0.27 trillion in 2000. The USA accounts for the largest share (about 79%) of the current total, followed by Europe (Kamaruzaman, Handrich & Sullivan 2010). By comparison, the Middle East and African region has a very small share (around 3%) (Kamaruzaman, Handrich & Sullivan, 2010).

The USA followed by the UK accounts for the world largest market for online retailing. Online retail in USA accounted for 3.6% ($142 billion) of total retail sales in 2008 (U.S. Census Bureau, 2010) and in UK accounted for 10.7% (almost $74 billion/52 billion EUR) of their retail trade in 2010 (Centre for Retail Research, 2010).

According to the Nielson (2010) report, the top 10 products/services globally sold online are books, clothing/accessories/shoes, airline ticket/reservations, electronic equipment, tours/hotel reservations, cosmetics/nutrition supplies, event tickets, computer hardware, videos/DVDs/games, and groceries.

It would seem that Saudi Arabia, as a leading world oil producer, would take advantage of e-commerce tools and applications. However, despite the fact that Saudi Arabia has the largest growing market for ICT products in the Arab region (Saudi Ministry of Commerce 2001; Alotaibi & Alzahrani, 2003; U.S. Commercial Services, 2008; Alfuraih 2008), e-commerce activities are not showing notable growth (Al-Otaibi & Al-Zahrani, 2003; Albadr, 2003; Aladwani, 2003; CITC, 2007; Agamdi, 2008). The Saudi Government's decision to introduce e-commerce in Saudi Arabia started in 2001. In response to the fast development of e-commerce around the world, the Saudi Ministry of Commerce established a permanent technical committee for e-commerce. However, this Committee has ceased to exist, and that the roles of e-commerce supervision and development were transferred to the Ministry of Communications and Information Technology (MCIT) in 2006. Since 2006, the efforts by MCIT to support e-commerce have not yielded results. When confronted with the lack of success thus far, MCIT simply claimed that they are still in the early stages of studying e-commerce (Ghawanny, 2011).

Unlike western nations, Saudi Arabian electronic systems have been concentrated on Saudi e-government development with e-commerce taking second place. In contrast, the development of e-commerce and indeed the Internet in Western countries has been driven by organizational and financial competition for market share – with the focus being on independent businesses. Typically, KSA and most Gulf states (Oman, UAE, Qatar, Kuwait and Bahrain) are late adopters of Internet technology, and while there is a rapid technological diffusion of internet technology there is a lag in the development of e-commerce adoption based on low level of public knowledge (and trust) about electronic models of business.

## 2.2. E-Commerce Maturity

Mature e-commerce is the state of having full development. Several models exist to help assessing the level of maturity in organizations. Maturity models include stages from an initial state to maturity to help organizations assess as-is situations, to guide improvement initiatives, and to control progress (Röglinger, Pöppelbuß & Becker, 2012). Morais, Gonçalves & Pires (2007) reviewed the principal models of e-business maturity and present the most cited models. Table 1 below demonstrates a comparison of the most cited maturity models.

**Table 1: Comparison of the Maturity Models**

| Model | Perspective | Development | Emphasis | Verification | Focus | Source | Stages |
|---|---|---|---|---|---|---|---|
| KPMG | Business | Linear | Non-specific | No | E-commerce | Private Sector | 3 |
| Model of Grant | Business | Linear | SME | Yes | E-business | Academia | 5 |
| Model of McKay | Technology | Linear | Non-Specific | No | E-business | Academia | 6 |
| Model of Earl | Business | Linear | Non-Specific | No | E-business | Academia | 6 |
| SOG-e | Technology | Linear | Non-Specific | Yes | E-business | Academia | 6 |
| Model of Rayport & Jaworski | Technology | Linear | Non-Specific | No | E-business | Academia | 4 |
| Model of Rao | Technology | Linear | Non-Specific | No | E-business | Academia | 4 |
| Model of Chan & Swatman | Business | Linear | Non-Specific | Yes | E-business | Academia | 4 |

Reproduced form (Morais, Gonçalves & Pires, 2007)

Among these models, Stage of Growth e-business (SOG-e) model is the more validated (Morais, Gonçalves & Pires, 2007) and accounts for both the Internet-based IT activity in organizations alongside with traditional information technologies and systems (Mckay, Marshall & Pranato, 2000). It was developed in 2000 by Mckay, Marshall & Pranato (2000) and verified in using empirical research assessing the progression of e-business maturity in Australian Small and Medum Enterprises (SMEs), (Prananto, McKay & Marshall, 2003).

The SOG-e consists of 6 stages (Mckay, Marshall & Pranato, 2000); no online presence, static online presence, interactive online presence, Internet commerce, integrated organization, and extended enterprise.

In stage 1 (no online presence), an organization has no clear direction for e-business. In stage 2 (static online presence), an organization is considering the importance of e-business; however, definite plan for moving forward does not exist. In stage 3 (interactive online presence), an organization is considering the significance of e-business and has definite plan for moving forward; however, the focus is on technology-centric perspective and not influenced by needs of business. In stage 4 (Internet commerce), an organization is making e-business adoption and development more business-focused, moving towards integrating and coordinating e-business components and the business processes.

In stage 5 (integrated organization), an organization Integrates between processes and activities of normal business and e-business. E-business initiatives, in this stage, provide strategic benefits through the building of strategic systems. In stage 6 (extended enterprise), an organization strongly integrates the e-business components and business processes within organization and with business partners making e-business involved with every aspect of the organization. E-business initiatives, in this stage, establish and hold strategic advantage. (Prananto, McKay & Marshall, 2003)

Each of SOG-e stages is assessed based on four layers: e-business strategy, e-business system, staff arrangement, and impact on business processes (Prananto, McKay & Marshall, 2003). The SOG-e is built on Galliers and Sutherland (G&S) model (1994) which has six stages to assess the maturity of traditional IS/IT within an organization (Mckay, Marshall & Pranato, 2000). In addition to the assessment of the maturity of traditional IS/IT within an organization, six stages are added/cooperated to assess the maturity of e-commerce and all together assess the maturity of e-business within an organization (Mckay, Marshall & Pranato, 2000). The SOG-e has, therefore, four stages to assess e-commerce maturity and four stages to assess traditional IS/IT maturity and the meeting point for these two stages groups is in stages five and six which serve to assess e-business maturity, see Figure 1 (Mckay, Marshall & Pranato, 2000).

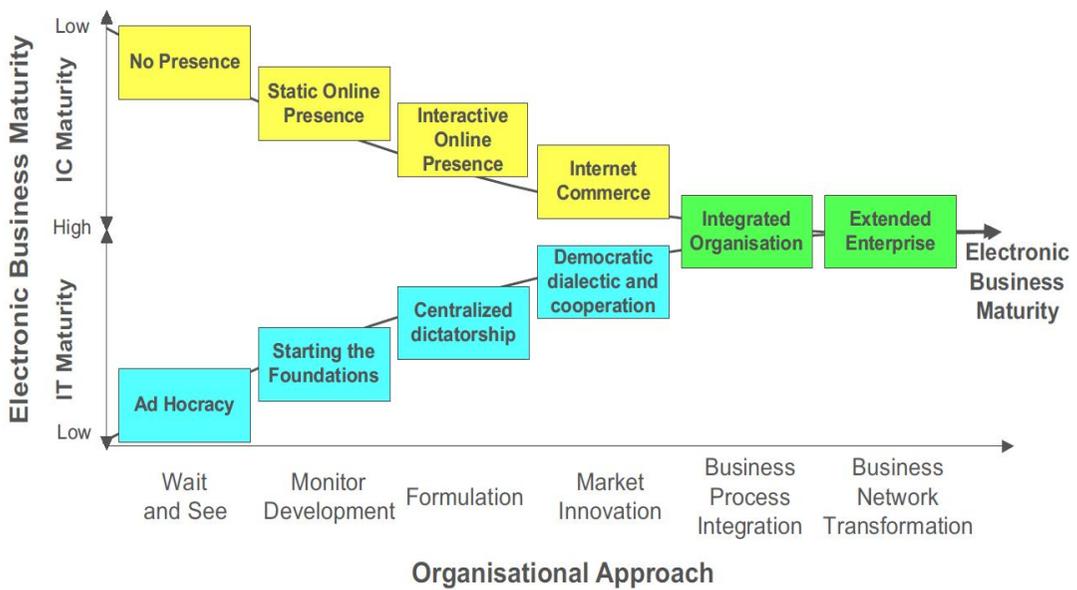

**Figure 1: The SOG-e Model of E-Business Maturity (Mckay, Marshall & Pranato, 2000)**

In this case, e-commerce is regarded as subset of e-business (Davis & Benamati, 2003) and this is the same definition which is the current study adopt. The current study adopt the definition of e-commerce that says e-commerce is commerce enabled by Internet including pre-sale and post-sale activities whereas e-business as supporting the range of business processes using all electronically mediated information exchanges, both inside and outside an organization (Chaffey, 2004). As the focus of this study is on e-commerce, therefore, the stages of SOG-e to assess e-commerce maturity will be adopted. Furthermore, due to the KSA is still in its early stage of e-commerce, only the first four stages of e-commerce maturity will be used.

## 3. Methodology

Since the research seeks to answer the question 'how', qualitative research seems more appropriate to guide the research (Creswell, 2013). Interviews were conducted with 16 retailers in Saudi Arabia. The selection took into consideration covering different types and sizes of retailing businesses. Sixteen interviews were carried out, which is an appropriate number for a qualitative study according to Creswell's (2013) guidelines.

The study's sample covered various categories of the retail market. Interviews were conducted with 16 retail decision-makers: 5 owners, 2 Chief Executive Officer (CEO), 2 regional directors, 5 directors, 1 marketing manager, and 1 IT manager. All were in Jeddah, the second largest and main economic city of Saudi Arabia. The sample covered 6 small, 5 medium, and 5 large organizations. The size of the businesses in Saudi Arabia are identified according to the number of employees; a micro size company has 1-5 employees, a small size company has 6-50 employees, a medium size 51-200 employees, and a large size more than 200 employees (Ministry of Commerce, 2009). The sample in the present study also covered different types of retail businesses: telecommunications, computers, sports, supermarkets, restaurants, printing services, Internet services, electrical and electronic products, beauty and body cares, books, watches and clocks, and chocolate and biscuit manufacturing. Nine out of sixteen of the participating businesses have a company website and only three of them sell online. Each organization's e-commerce maturity stage was determined based on whether no website existed (stage 1), a static website existed (stage 2), the existence of an interactive website (stage 3), or the existence of an e-commerce website which allows consumers to place their orders and pay online (stage 4). In terms of e-commerce maturity, three retailers were classified in stage 4, two retailers in stage 3, five retailers in stage 2, and 6 retailers in stage 1. The classification criteria for e-commerce maturity model were based on whether they have or do not have existence of static, interactive or e-commerce website.

The qualitative data obtained from the interviews was analysed using content analysis. Content analysis is a research technique for studying texts, breaking it into small component units and finding relationships between these units to give a meaning (Denscombe, 2010). In inductive content analysis (Elo and Kyngäs 2008) a researcher reads the content and defines the codes as they emerge during data analysis; that is, the codes are derived from the data (Hsieh & Shannon, 2005). The current study used inductive content analysis because the study did not aim to test pre-identified concepts but rather used a more exploratory approach to investigate issues influencing retailers on whether to adopt or not adopt e-commerce.

After preparing data, the coding process started. The process of coding involved combining the text into small categories of information, and then assigning a label to the code. The coding process can be defined as "aggregating the text or visual data into small categories of information, seeking evidence for the code from different databases being used in a study, and then assigning a label to the code" (Creswell 2013, p. 184). In the coding process, transcribed interviews were read line by line looking for segments that can give meaning and these were labeled with codes. The outcome of this process provides a long list of codes, in the present case 82 codes. Another step was required to reduce the number of these codes by combining similar codes together. This step is referred to as 'focused coding'. The purpose of focused coding was not only to reduce the large number of codes by combining the similar codes, but also to produce a structured list of codes.

Re-reading of the texts and consideration of all 82 codes in the outcomes of this process led to six categories encompassing 55 codes: consumer issues, organizational issues, environmental issues, market issues, retailer perceptions, and general issues.

Elo & Kyngäs (2008) suggest using a panel of experts to support the content analysis outcomes of interviews. Byrne (2001) also suggests that using people other than the primary researcher to assess the findings is a good strategy for conducting credible qualitative research. Thus, an independent academic colleague went through all the interview transcripts and assessed the coding. Given the feedback, to organize the codes under three main categories: consumers related factors, environment related factors, and organization related factors; all the codes were then organized under these three categories. For further qualitative analysis quality assurance, all the code statements were organized under the three main categories together with examples of statements by the participants that the researcher believed exemplified the code statement and were shown to another academic. This led to 22 code statements. These code statements can be considered suggested factors that influence retailers' decision on whether to adopt or not to adopt e-commerce.

## 4. Analysis and Discussion of the Interview Data

Table 2 presents the codes statements resulting from the interviews data analysis which can be called factors influencing retailers' decisions to adopt e-commerce in Saudi Arabia.

**Table 2: factors Influencing Retailers' Decisions to Adopt E-Commerce in Saudi Arabia**

| Consumers related factors |
|---|
| • Cultural attitudes or habits regarding online shopping |
| • Consumers' familiarity with e-commerce |
| • Consumers' understanding of e-commerce benefits |
| • Consumers' trust |
| • Consumers' reluctance to use credit cards |
| • Consumers' level of demand of buying online |
| • Consumers' knowledge to buy online |
| • Consumers' willingness to pay for delivery fees |
| Environment related factors |
| • Internet users in Saudi Arabia |
| • Wi-fi & broadband services availability |
| • Protection system |
| • Required action by government |
| • Online payment systems |
| • SADAD, national online payment system |
| • Issues relate to the Saudi mailing and addressing system |
| Organization related factors |
| • E-commerce difficulty |
| • Issues related to the nature of products |
| • Business familiarity and knowledge with e-commerce |
| • Management Attitude toward e-commerce |
| • Business priority |
| • Security and trust concerns |
| • Setup and maintenance cost concern |

The discussion will be grouped around the 22 factors developed as explained in the methodology section and will also highlight comments of retailers in companies at different stages of e-commerce maturity. It is interesting to see how retailers in companies of different stages of e-commerce maturity might differ in their attitudes regarding the 22 factors. Where comments of interviewees are given, the pseudonym is given, together with a number in brackets which indicates the e-commerce maturity stage of the participant's company.

As discussed earlier in the earlier methodology chapter, the meaning of these numbers are as follow: stage 1 indicates that a company has no online presence/ no website; stage 2 indicates that a company has a static website that gives some information about a company; stage 3 indicates that a company has an interactive website that accepts feedback, comments, and communication from consumers; stage 4 indicates that a company has an e-commerce website that allows consumers to place their orders and that accepts their payment online.

4.1. Consumers' Factors/Issues/Concerns

Interestingly, the retailers' interviews raised several issues relating to consumers that could discourage them from adopting e-commerce in Saudi Arabia. These issues include the retailers' perceptions of: cultural attitudes or habits regarding online shopping; consumers' familiarity with e-commerce; consumers' understanding of e-commerce benefits; consumers' trust in online shopping; consumers' reluctance to use credit cards; consumers' level of demand of buying online; consumers' knowledge of how to buy online; and consumers' willingness to pay for delivery fees.

4.1.1. Cultural Attitudes or Habits Regarding Online Shopping

From the statements made by interviewees it appears that some retailers believe that the culture of the people in Saudi Arabia to buy or sell online is such that it discourages retailer to adopt and online retailing system. The retailers believe that people in the KSA prefer traditional shopping, visiting the shops, and inspecting the quality and uses of the products before making a decision to buy.

Saeed (1), Saleh (1) and Waleed (1) all stated that online shopping has not been a habit for Saudis and that they would not benefit from such selling online. Their comments are all close to what Saeed (1) says: "*the culture of people is not encouraging, so it is not useful for us*".

Participants whose companies only have a static website did not seem different from participants with companies at stage 1. They raised the same issues that because consumers in Saudi Arabia do not have a habit of online shopping they do not sell online.

Ahmed (2) and Nasir (2) made statements close to what Talal (2) said: "the selling with customers online is not useful for us due to the habit of people. When the culture of people changed and accept to buy online we will apply e-commerce system".

Their thinking is all about whether they will gain more profit from selling online. From their perspective they think that e-commerce is not useful for them unless consumers accept it and it becomes a habit for them.

What about companies that do sell online in Saudi Arabia; do they think about consumers in KSA and their shopping habits? Two participants, Salem (4) and Thamer (4), agreed that online sales cannot compare to normal sales in KSA. They stated that they know that selling online in Saudi Arabia does not give more profit. However, their companies want to be active and take a positive role or to experiment in the new market. The main reason motivating them to sell online is to open up a new marketing channel and establish a good reputation for their companies' e-commerce in the region. Some believe that e-commerce in Saudi Arabia has a bright future. Thamer (4) stated that "*I believe that the Saudi market is bullish and a fertile ground for investments, but –in my opinion- even if we start thinking about e-commerce we need time, not less than 10 years, to reach maturity in this field*".

### 4.1.2. Consumers' Familiarity with E-Commerce

Interestingly, retailers who raised the issue that consumers are not familiar with e-commerce have no experience selling online. Ali (1) stated that "*the familiarity, sellers and customers are not familiar with online sales and purchases*". This retailer has no experience buying online, does not know friends who have experience buying online, and his business is not involved with e-commerce at all. So, Ali's thoughts about consumers are based on his surrounding environment, and are not based on practice.

On the other hand, Muhammed (3) seems more optimistic, saying "once the businessmen or the ones who run the business in KSA get the confident that the people are more keen to go online and visit the places they will be more than happy to jump to the bandwagon and put their business in e-commerce channel". Muhammed's company, which sells electronics and home appliances, is planning to sell online in the near future. They have almost everything ready to go ahead. So it seems that his impression about his consumers' familiarity with e-commerce is positive.

The situation here is like a chicken and egg dilemma; which should start first? Should companies start selling online and hope the consumers will follow or should companies wait until consumers show stronger movement towards online shopping. Osam (3) is of the opinion that it is best to wait until people become familiar with online shopping, stating that "*in general people have to become familiar with this system before we sell online*". Similarly with Talal (2), "*when people accept to buy online we will apply e-commerce system*". By contrast, Thamer (4) and Salem (4) did not agree to wait for consumers to be mature in purchasing online. They knew that they will not gain more profits or many online consumers similar to online retailers in western countries; "*we already know that this region doesn't give you more sales on the Internet because that habit of the people here in this region used to go outside and buy*", Salem (4) stated. However, they "*continue providing this option* [selling online] *to encourage people to use it*", Thamer (4) said.

4.1.3. Consumers' Understanding of E-Commerce Benefits

Two participants raised the issue of consumers' understanding of e-commerce benefits and they are totally different in their opinions. Talal (2) saw it negatively, that consumers in Saudi Arabia do not understand the benefit of e-commerce. "*If there are people who understand the benefit of e-commerce I agree with you it help to gain more profits*". . For this reason, it seems Talal believes that there is no point in using e-commerce becaue people do not understand its benefits and would not buy from business online. Mohammed (3) totally disagrees with that judgment about consumers.   He said: "People here understand the benefit of e-commerce 100%... I've seen people there are the indigenous original Saudi people superbly well educated and they know exactly the benefit of e-commerce and everything".

Again, personal feeling or experience seems to play a role here in these two different opinions. The first participant, Talal (2), has no experience purchasing online and his company has no engagement in e-commerce activities. Therefore, his judgment is not based on practice. By contrast, Mohammed (3) has lived in western countries and has engaged in buying online many times.

In addition, Mohammed's company has an interactive website gaining consumers' feedback and comments regarding their products. His company has done almost everything to move forward for the next step and to sell online.

### 4.1.4. Consumers' Trust

The issue of consumers' trust has received high attention from the retailers who were interviewed. Interviewees often raised the issue that they believe that consumers are sensitive and uncomfortable about online shopping because it involves paying online and there is an absence of inspecting the product in one's hands. While participants whose companies are classified in level 1 and 2 in e-commerce maturity raised their concerns that there is no trust in online shopping; participants whose companies' e-commerce maturity levels are 3 and 4 seem to have moved beyond raising the concerns and instead suggest solutions to overcome this issue and build trust with online shoppers as it is explained in the following passages.

Tameem (1) clearly stated that he does not trust buying online from Arabian businesses. Later when he was asked why his business was not taking steps to sell online, he replied: "*if people knew that the one who run an e-commerce website is Arabian or the company owned by Arabian, they will be unmotivated to buy*"! Similarly in Nasir's (2) case, he has negative feelings regarding dealing with money over the Internet, even with using Internet banking he is sensitive. The interesting thing here is to know that Nasir's company accepts orders using phone calls or Fax and receives payment from consumers using bank transfer, and ships ordered products with well known international shipment companies such as DHL and FedEx. Nasir (2) commented regarding selling online that "*there are people when they pay online remain in doubt and not sure of receiving their purchased orders*". Interestingly, he thinks his consumers remain in doubt when they purchase online and but that do not have doubts when they order using the phone or when they transfer money using a bank account.

Moving beyond raising the concerns of consumers trust, other participating retailers suggested solutions to overcome this issue and build trust with online shoppers. Most of the participating retailers who suggested these solutions are involved in e-commerce activities.

Most of the suggestions to overcome consumers trust issues were concerned with providing alternative ways of paying online to help customers to choose the one they feel safe with. It seems that some retailers believe there consumers in Saudi Arabia have an exaggerated of using their credit cards online as Osam (3) commented, "*people are reluctant to use credit cards, I think that this fear is exaggerated*". The national payment system SADAD was mentioned as a good solution here to gain consumers' trust.

It is used for online purchases in Moneer's (4) company and he commented that "*SADAD is great idea and more secure than credit cards and encourage people to buy online*". The PayPal system for paying online was also mentioned as a method that helps gain trust with online consumers. Naif (2) commented regarding the process of PayPal saying: "*with this procedure customers feel happy to deal with this intermediary e-payment option which protects their rights and also build the trust with companies that deal with PayPal*". The online payment systems are further discussed in the environment related factors section.

Other solutions that may work in Saudi Arabia to build trust with online consumers were the involvement of the government and existence of e-commerce protection system. Thamer (4) suggested that "*citizens will have more trust if this subject sponsored by the government because we, in Saudi Arabia, have great confidence in anything that comes through the government*". Salem's (4) company is "*working to have certificates from trusted organizations to build the trust with our customers*". Salem (4) suggested that "*it should be there is a certification body from the government... this is good to build the customers trust with the certified companies as the government trust them*". Ahmed (2) also suggested that the government build clear legislation system for e-commerce. Ahmed (2) stated that "*customers have to be ensured there are rules and legislations protect their right. This is very important to build the trust with customers... I advice the main supporter in this field is to find out a way to build the honesty/trust between customers and sellers*". These two issues of government involvement and are further discussed in the environment related factors section.

4.1.5. Consumers' Reluctance to use Credit Cards

Interestingly, some participants who raised the issue of their belief that consumers are reluctant to use credit cards for online payments do sell online. Their statements were based on experience.

For example for Salem's (4) company they notice that their consumers feel difficult to put their credit cards information on their e-commerce website, so they may contact them for arranging other options for payment as he stated *"we notice that people feel difficult to put their credit card information… we have another way here especially for local customers, we have call centre and the customer can call us (free calls) and make the order by phone and go to pick it up from the branch"*.

Salem's company is working to provide more options to pay for online purchased ordered to overcome the issue of consumers' reluctance of using credit cards as he considered the importance of *"providing more online payments because not everybody has credit card, it should be there is another ways of online payments"*.

Similarly with Moneer's (4) and Thamer's (4), they partly refer the low acceptance of online shopping among Saudis to the consumers' reluctance to use credit cards. Thamer (4) stated that consumers *"are afraid to put their credit cards number on the internet"*. And he suggested *"for local banks to offer other easy options for online payment to encourage people to purchase online"*. Providing other options for consumers to pay online seems very important for these retailers. For example, Thamer's company provide the option of Cash on delivery. Cash on delivery seems acceptable for Thamer's company because the company has its own delivery system and purchased products delivered by the company's staff using its own trucks. Other option for online payment that has become popular and trusted for Saudis to pay for e-government services is the national payment system SADAD. Recently, SADAD has become available for commercial businesses and private companies to use allowing their consumers to pay online. Moneer's (4) company provide SADAD for their consumers to pay for their online purchases, *"we provide two options, credit cards and SADAD system"*, Monner  siad. Moneer (4) commented on using SADAD in his company that *"SADAD system is more secure system and great. With this system there is no need to entre personal payment details on our website"*. These two empirical examples suggest that for companies in Saudi Arabia, using more options for online payment is critical to overcome the issue of consumers' reluctance using credit cards.

### 4.1.6. Consumers' Level of Demand of Buying Online

Participating businesses considered consumers purchasing power in Saudi Arabia as weak and not supporting them to run their businesses online. The issue of low level of demand raised by participants in all different four groups of e-commerce maturity levels. Saleh (1) commented that "*the current situation here is not encouraging us to go ahead with this idea*" and Ahmed (2) stated that they "*have tried to market on the Internet but there is no demand*". While it is not based on practice to hear from businesses which are not involved in e-commerce activities, i.e. Saeed (1), Saleh (1) and Ahmed (2), that the level of online purchase demand is low; this was confirmed by businesses which are involved in e-commerce activities. Thamer (4) and Salem (4) confirmed that the online sales cannot be compared to the normal way of selling. Very few consumers choose to buy from their businesses online as they stated.

Thamer (4) commented "there is no good demand to buy online. For example, in western region of Saudi Arabia we receive 3-5 orders buying online every week and these are nothing compared to normal way of business". Similarly with Salem (4) as he stated that "we are not in Europe or America where people easily buy from the Internet, culturally we are different than them that's why e-commerce is not getting much more businesses".

It is again about the chicken and egg dilemma, which should start first companies sell online or consumers buy online. While many businesses seem they adopt wait to see strategy, there are other businesses despite they mentioned the issue of low level of online purchases in Saudi Arabia, they choose to be among those explorers, motivators and developers of e-commerce in the region. For example, while Thamer (4) and Salem (4) acknowledge the low level of consumers online purchases demand as Thamer (4) stated "*there is no good demand to buy online*", they choose to go ahead and "*continue selling online to encourage people to buy online*", Thamer (4) said. It is not only that the main reason motivating them to sell online but rather other factors contributed to this. For example, these companies have well established e-readiness and sell online was considered to them as additional marketing channel. Furthermore, they want to have strong foothold in e-commerce in Saudi Arabia.

They see that e-commerce has promising future in the region, "*I believe that Saudi market is bullish and a fertile ground for investments, but –in my opinion- even we start thinking about e-commerce we need time not less than 10 years to reach the maturity in this field*", Thamer (4) said. It seems correct that the level of online purchase demand in Saudi Arabia is low compared to western countries; however, Saudi Arabia has huge potential for e-commerce market (Gabr, 2013). This issue is further discussed in the discussion chapter.

4.1.7. Consumers' Knowledge to buy Online

Interestingly, participants who have perceptions or raised the issue of consumers' knowledge to buy online are the ones whose companies do not sell online, and interact online with consumers. Their given statements are like prejudgments that consumers do not have the knowledge to buy online! Similar to what have been said earlier in consumers' familiarity and benefits understanding of e-commerce sections, again personal feeling or experience plays a significant role here in these perceptions. It is not logical to judge the consumers' knowledge to buy online while you have not interacted with them online!

Taking Ali's (1) statement as an example, "*the majority of people in our country do not know how to buy and sell on the Internet*". Including the word 'sell' in his statement is a key to what he meant. He included consumers being not knowledgeable to buy online as same as he/his company was not knowledgeable to sell online.

Similarly with Nasir (2) as he commented that "*there is ignorance in the community. Most of them don't know the meaning of e-commerce*". It seems and can be understood from these generalised statements they based on their own/ or their company knowledge and experience. For example, Nasir (2) stated that "*honestly there is ignorance inside our company in terms of e-commerce*". A question may be asked here, how for a person judge others knowledge about something he/she is not knowledgeable about it? In addition, none of these participants, (i.e. Ali (1), Saleh (1), Nasir (2), and Talal (2)), has experience buying online, they never purchased online. They also answered that they have no friends or relatives who do have experience purchasing online.

While it is not logical to accept this judgments, it still worthy to consider this perception that they have about consumers as significant factor influencing their decision to use e-commerce. Education and building awareness programs may be useful in this regard. Sellers may need to be aware of consumers behaviours buying online based on research reports and actual financial spending. This may add value changing this negative perception about consumers and as result helps to positively influence their decision toward using e-commerce.

4.1.8. Consumers' Willingness to Pay for Delivery Fees

The same as the previous issue, participants who have perceptions or raised the issue of consumers' willingness to pay for delivery fees are the ones whose company do not sell online, and interact online with consumers. Their given statements in this regards are based on as they perceived not on empirical evidence. Again, it seems they have this perception based on their own experiences. None of these participants, Saleh (1) and Hassan (1) except Ahmed (2), has experience buying online. Ahmed (2) only has bought one time an airline ticket.

I can see also one of the inhibitors is delivery fees, Ahmed (2)

We have concern regarding delivering the purchased products and its fees, Hassan (1

If you request from the customer to pay extra 10 SAR for such a fees like this and if the price in total comes more than the price in the local shops, he/she may buy it from them without buy it from you online, Saleh (1)

By contrast, businesses that involved in selling online had no concern regarding the delivery fees. This is because they have known and gained experiences on how to deal with delivery companies and reduce the cost of shipments. For example, Salem's (4) company has had contracts with FedEx well known and trusted international shipment company, "*we have done is having very strong contract with express and trusted shipment company, FedEx*", Salem (4) said. They ship for them for very competitive prices. Salem also mentioned that if a consumer or his employee has a contract with any shipment company, a consumer is able to provide them with his/her shipment company reference number when completing the order online, so the order is delivered with no cost.

Salem (4) commented *"we have also another shipment type, the customers who have accounts with the shipment companies they can provide us their account numbers with anyone of these companies and we will arrange with these companies to come and pick up your products and they will charge you as the same agreement you have with them"*. These are examples demonstrate how these companies came up with solutions to reduce the cost.

However, reducing the cost and looking for competitive prices are not always what consumers looking for. This is what retailers in the lower stages of e-commerce maturity not aware of. In many cases the convenience is about having the product delivered to a customer's home, at no extra cost, so that she/he does not have to transport it and install it her/himself. It is about not having to find a car park. It is about not having to have a car. It is about placing an order if the goods are not in stock so that they are delivered as soon as they are available. It is about being able to compare the same goods (or comparable goods) from a number of online suppliers whilst on the computer rather than walking from shop to shop. By offering comparably lower prices, consumers would have the choice of either spending less money, or purchasing a higher quality item online, for a price comparable to a lesser item in a bricks and mortar establishment.

## 4.2. Environment Factors/Issues/Concerns

Environmental factors refer to the environment of technological readiness and other involved factors to serve the running of e-commerce.

The literature discusses various environmental issues that influence rate of e-commerce growth in a particular environment. The raised factors, issues or concerns by the participants regarding the e-commerce environment in Saudi Arabia include Internet services and users, e-commerce protection system, required action by government, online payment system, national payment system, and the Saudi mailing and addressing systems. Here is the discussion of all these issues with sentences being quoted from the retailers' comments to support the discussion.

**4.2.1. Internet users in Saudi Arabia**

There is no doubt that the low percentage of internet users in an environment is an inhibitor to the growth of e-commerce. On the contrary, high percentage of internet users plays significant role in the development of online services and attracts more businesses to be online. The internet users in Saudi Arabia are seen high by some participating businesses and may play a motivational role for e-commerce success. "*Today almost every home has access to the Internet, many people browse it and of course this brings customers to your shop*", Hassan (1) said.

As discussed earlier in the background chapter, the internet users in Saudi Arabia is increasing rapidly. The Saudi market has becomes attractive for more online retailers in the Arab world (Gabr 2013). To what extent this would play a motivational role for Saudi businesses to sell online? Despite the high numbers of Internet users in Saudi Arabia, some participants recognised the activities of e-commerce are very low but acknowledge that will be changed in the near future and e-commerce in Saudi Arabia has promising future. Thamer (4) believed that "*the Saudi market is bullish and a fertile ground for investments, but –in my opinion- even if we start thinking about e-commerce we need time, not less than 10 years, to reach maturity in this field*".

4.2.2. Wi-Fi & Broadband Services Availability

The e-commerce literature places great emphasis on the wide availability of broadband Internet as the backbone of ICT infrastructure strength. Despite the recent development in broadband services compared to previous years in Saudi Arabia, still more efforts are needed as suggested by some retailers. The CEO of a large-size company that runs a business selling complete home and electronic solutions, Mohammed (3), see that the current situation of Internet services and access in Saudi Arabia needs to be improved in order to create more attractive environment for e-commerce.

Mohammed (3) commented "*I think someone needs to kick the whole things of having the access of Wi-Fi, of having the access of broadband, having the access in every house hold. Once this done, obviously you will see the results of that very soon*".

4.2.3. Protection system

Regulations and rules, in commercial transactions, are very important in terms of systemizing the work, and protecting the rights for all involved parties.

Some retailers, in the current study, urge for e-commerce consumers protection system. For example, Ahmed (2) saw that "*there is no clear system to protect sellers and customers rights… the system that protects customer right when he/she pays, he/she will receive the purchased products/services must be ensured*". Interesting is to hear that businesses urge for consumers protection system. It is understandable why these retailers urge for having legislation system for e-commerce. With the lack of legislative system for e-commerce, trust is difficult to be built with consumers. Businesses which raised this issue they stated that setting up legislative system is required action by government. As discussed in earlier section, consumers trust this type of procedures of having protection system and enhanced by the government helps to build trust with consumers. This is confirmed by Thamer (4) and Salem (4) whose companies already involved in selling online. They have not clearly indicated that a clear e-commerce law should be exist but they mention the involvement of the government as a key to develop trust with consumers. Their views were discussed in the next section.

4.2.4. Required Action by Government

Government support for e-commerce growth takes various forms from country to country such as technical support, training, funding provided ensuring coherent policy for IT and e-commerce for consumer protection; secure, transparent, predictable, and enabled environment; support, coordination, collaboration, and cooperation.

As discussed in the earlier section, businesses needs government to be involved in e-commerce as a third party which helps businesses to build trust with consumers. The businesses which urge for government involvement were the ones which they are sell online. Salem (4) and Thamer (4) saw that government involvement helps them to build trust with their consumers.

Salem (4) suggested that "*it should be there is a certification body from the government itself to say that this company is a certified company by local government and you can buy from them. This is good to build the customers trust with the certified companies as the government trust them*". Similarly with Thamer (4), it is all about contributing to building trust with consumers, "*citizens will have more trust if this subject sponsored by the government because we, in Saudi Arabia, have great confidence in anything that comes through the government*".

4.2.5. Online Payment Systems

Providing e-commerce services with the necessary financially secure transactions requires trustworthy online payment mechanisms. Participating businesses in the current study which raised the issue of online payment systems were all agreed that providing more options making them available for consumers to choose from the option that they feel safe with is significant incentive for consumers. On the businesses side, the wide available online payment options for businesses to implement for their e-commerce systems, the higher they accept to sell online. It is again the matter of providing consumers with means that contributes to build trust with businesses. For that reason, Thamer (4) emphasis on providing local payment systems that helps consumers to get used to use. Credit cards are fine and widely use world-wide, however in Saudi Arabia credit cards seem yet not popular. Thamer (4) suggested that "*for local banks to offer other easy options for online payment to encourage people to purchase online*", similarly with Salem (4) and Osam (3). Osam (3) suggested that "*banks should provide easy options to have two credit cards, one with large amount of money and another one with small amount, called debit cards, to be used in online payment*". This suggestion is good; however, local banks in Saudi Arabia charges fees for having debit cards or intrant purchases cards and that is why there is a need for improving this situation.

While some businesses urge for having local solutions for online payments which helps to build trust with consumers and encourage them to buy online, there are some other businesses that use international payment system such as PayPal. However, PayPal still not popular payment method to be used in Saudi Arabia. One popular method that is widely used in Saudi Arabia for paying bills online is SADAD. This system seems to be very good solution for providing trustworthy online payment method that is accepted by consumers. SADAD is discussed in the next section.

4.2.6. SADAD, National Online Payment System

Following from the previous point that businesses are looking for online payment systems that help to build trust with their online consumers. The national payment system SADAD seems great for providing mean to pay online and build trust with online consumers. SADAD is a national electronic bill presentment and payment service provider for the Kingdom of Saudi Arabia, and was launched on October 3rd, 2004. The core mandate for SADAD is to facilitate and streamline bill payment transactions for consumers through all channels of the Kingdom's Banks.

It relies on existing banking channels (such as Internet banking, telephone banking, ATM transactions and even counter transactions) to allow bill payers to view and pay their bills via their banks. It has been found that consumers are comfortable and tend to have more trust with using SADAD in e-government services.

SADAD may represent good solution to overcome the issue of consumers trust paying online. In the current study, a personal experience of one participant purchasing online using SADAD has leaded him to influence his company providing SADAD for their consumers as an option for online payment. Moneer's (4) company provide two options for its consumers to pay online: credit cards and SADAD system. Moneer (4) commented on using SADAD in his organization to complete online purchase orders saying "*SADAD is great idea and more secure than credit cards and encourage people to buy online*". Moneer (4)

At the time of conducting this study, SADAD was limited for use to 100 billers, mostly large and government organizations. For that reason, SADAD was not considered as online payment method for e-commerce. Salem (4) commented "*SADAD before was very expensive solution, it is good for large companies but it's not for middle size company like us*". However, recently this limitation have been addressed and the government working on billers base expansion, which will increase its biller options from 100 to 20,000 (CITC, 2011).

4.2.7. Issues Relate to the Saudi Mailing and Addressing System

There are some retailers in Saudi Arabia that have their own delivery system. There are some others who have no problem with the delivery of phone orders.
They organize with local shipment companies to deliver the goods, however, this is not considered as a professional delivery service because these normal delivery companies do not get a clear home address, so they deliver to their offices only and then contact the customer to come pick up his/her order. When a customer wants the products to be delivered to their home, they have to arrange this with the delivery company and pay an extra fee. They have to explain to the driver where their home is located.

With the recent development in the mailing and addressing systems in Saudi Arabia that each building has address and mailbox, still some companies work using the old system relying on phone numbers. For example, Salem's (4) company "*depend on the mobile phone number; if the mobile phone number is not correct then it's difficult to deliver the product*". It seems part of the problem with the international delivery companies that they do not follow the update and new addressing system in Saudi Arabia which carried by the Saudi post. The Saudi post delivers to homes. In addition, Some businesses have lack of knowledge of the efforts made by the Saudi Post since 2005. For example, Salem (4) thought that "*there are certain locations (e.g. companies' buildings) you can know the address clearly, you know which street and block, building, floor, flat etc; but with community houses this is not very clear*". Similarly with Osam (3) where he commented on this issue saying "*the problem is mail address; still some people do not have mail address*".

There have been recent developments in the mailing and addressing systems carried out by the Saudi Post. Huge efforts have been done naming streets and numbering all residential subdivision inside cities in Saudi Arabia. One possible factor, pertaining to the low use of building addresses, might be the lack of awareness either by retailers or consumers of these recent changes in the Saudi mailing and addressing systems.

### 4.3. Organization Factors/Issues/Concerns

The participating businesses has raised several issues relate to their organizations that affect their decision to adopt e-commerce. These issues include e-commerce level of difficulty; issues related to the nature of products; management attitude toward e-commerce; e-commerce familiarity and knowledge; business priority; security and trust concerns; setup and maintenance costs. Below is the discussion of all these issues with sentences being quoted from the retailers' comments to support the discussion.

### 4.3.1. E-Commerce Difficulty

The difficult tasks of e-commerce are different based on each company e-commerce maturity level. For example, participants whose companies in level 1 and 2 of e-commerce maturity model raised issues regarding creating e-commerce website, train employees, online payment and products delivery. For example, Saeed (1) commented that "*e-commerce is difficult; you have to design a commercial website, train your employees, deliver goods, etc*".

The raised concerns regarding creating e-commerce website, online payment and products delivery are not seen difficult by the participants whose companies already sell online. Of course this is because the strong e-readiness of these companies. However, part of their success is their knowledge and experience to find solutions. For example, creating e-commerce website is not seen an issue by Salem (4). He commented that *"building the website is very easy. You can buy cheap stuff/software from the Internet, there are e-commerce packages available cost around $100 to buy a package and publish your e-commerce website"*. Similarly with Thamer (4), *"there are complete e-commerce solutions that you can easily use"*. It is the same thing that was discussed in paying delivery fees issue that somebody may have negative perception bout something and consider it difficult, expensive or not useful while that in reality is not right. The issue here is about knowledge, lack of knowledge cause a person to pre-judge.

Other issues regarding the difficulty of e-commerce relate to the market when sell online. These issues are all made around the difficulty to compete online while products are available in the local market. Mostly, businesses in level 1 and 2 raised the difficulty to compete online while products are available in the local market. They consider e-commerce is not useful for products that can be found in the local market, e-commerce is useful for products that is unique and does not exist in the local market. Here are examples of what these businesses thought.

I agree to go ahead with this idea if there is products are not available in the local market where we can have competitive advantage. Ali (1)

Computer shops are widespread in Jeddah, so to go ahead with this idea is not benefiting us, Saleh (1)

Having no similar products in the local market help to achieve high profits, but if the products are available in the market, I believe that profits will be very weak. Waleed (1)

The raised issue regarding the difficulty to compete online is an interesting and important point for discussion. The usual way around this is to think of different online business models, with efficiencies to reduce the cost of doing business. It is much like KSA moving from petrol to an information economy as retailers must reduce the reliance on transport (cost of petrol and time) and storage of stock (rental).

For example, Amazon can sell books cheaper than many local bookstores because it does not have to rent shop space or run a delivery service itself. Order details are sent directly to publishing houses for direct delivery by a specialist courier or postal service. There are also models that provide certain items only by Internet sales so customers have to use that channel if they want that product. Similarly, access to specialist items direct from manufacturers or wholesalers is a price reduction model. Part of the problem may be that Saudi retailers are not aware of the different online business models, and that they can take advantage of different forms of disintermediation – removing service costs from the retail price. This lack of education is part of the problem surrounding e-commerce – not just for retailers, but for the customers as well. Saudis are not adept at creating a competitive advantage online yet. It is a new way of thinking about supply chains.

4.3.2. Issues Related to the Nature of Products

According to the statements made by interviewees, it appears that the decision of whether or not to adopt an online retailing system depends on the type of business or product. For example, companies that sell food and fragile products are reluctant to sell online due to their concerns that the goods may not be received in the same condition in which they were shipped.

Some of these issues relate to the ability of businesses to deliver or know delivery company that deliver in satisfactory situation. For example, Saeed's (1) business involves with selling grocery. He commented on delivery saying that "*delivering goods need special care with some of our products because they require being stored in a specific temperature to delivered safely and healthy*". That is the case that Saeed's (1) company does not have the ability to have refrigerated trucks to deliver their product and do not know delivery company to do so. It is not like other products that do not need special care.

However, Salem's (4) company sell chocolates and biscuits which need to be stored in cool area. Although their products fragile and need special cares, they manage well to deliver their products in satisfactory conditions. Despite Salem (4) raised concern because customers are afraid that the fragile product may not be delivered in one piece, his company is managing well with FedEx, delivery company, and have had no complains form their consumers.

Salem (4) commented "*some customers are interested to buy from you but when they see fragile products they will go away, it is better for them to buy from local market instead of buying online where they cannot ensured the product will shipped safely. To reduce this thing what we have done is having very strong contract with express and trusted shipment company, FedEx*". Therefore, it is again about knowledge and gaining experience to know more options and alternative.

Furthermore, delivery of special products is not the only area of consumer concern as indicated by retailers; there are other issues involved with the type of product e.g. a beauty company selling: perfumes, makeup, cosmetics and body lotions, shampoos and skin care. Ahmed's (2) business involves selling beauty products. He commented that it is difficult for them to sell online, consumers will not accept it because "*our products related to smell, shape which need to be physically seen*". While it is agreed that some consumers need smell, touch and inspect products in their hands, many successful businesses exist in selling beauty products online, e.g. the Body Shop. It is about consumers get used to know products. The products that have high demand purchased from normal shop, they are the ones have more chance to be successful online. This is because consumers they already know that product, know smell, shape etc and they will not have any problem buying it online once the trust is established. Again trust is mentioned here. It is more likely to be trust issue not the type of products issue.

### 4.3.3. Management Attitude Toward E-Commerce

The management attitude towards the changes is a key for organizations to adopt e-commerce. Rogers (2003) identified the attitude of an organization's management towards change as critical factor in terms of an organizational innovativeness.

The current study highlights positive and negative attitudes of management toward e-commerce and how they influence the adoption decision.

Mohammed (3) has positive attitude toward e-commerce and has well experience using e-commerce as buyer. He is "*personally like to do move toward e-commerce system where customers will be able to browse, search, check the prices for competitor, and check the quality of products in terms of features, advantages and benefits and then make the right choice. It makes comfort and less time to move around*".

This positive attitude led to that Mohammed's company is almost ready to start selling online. Their website currently display products, features and prices, and gains feedback and interacts with consumers on their website.

Another positive attitude for another case contributes to the continuance of selling online despite the low level of online purchase demand is with Thamer's (4) case. Despite his evaluation that consumers purchases online is very low, his company continue offer online channel for its consumers because they that the Saudi e-commerce market is bullish and a fertile ground for investments. Thamer (4) commented "*I believe that Saudi market is bullish and a fertile ground for investments… we will continue providing this option* [selling online] *to encourage people to use it*".

In contrast, negative attitude towards e-commerce influence negatively to the adoption decision. Despite Nasir's (2) company is accept orders by phones, payments by bank transfer and deliver products with DHL and FedEx, they do not accept this process to be complete on their website due to this negative attitude. The management attitude in this company have negative attitude toward using Internet to sell and buy and pay online. They think that "*there are thefts of credit cards numbers, and there are hackers penetrate your privacy. So this is a problem for the money dealing on the Internet*", Nasir (2) said. In this case it is clear that the negative attitude makes no room for thinking differently or looks for alternatives. Since Nasir's company has their website demonstrating the products descriptions and prices, accept orders by phones, payments by bank transfer and deliver products with DHL and FedEx, it seems not an issue to accept orders on website and accept the payment as they wished, i.e. bank transfer. However, this negative attitude makes no room for thinking differently.

4.3.4. Business Familiarity and Knowledge with e-Commerce

It's normal to find resistance to the acceptance of a new technology or idea, as people need time to become familiar with. Regers (2003) showed an adoption timeline for a new technology/idea where there were few adopters in the beginning, and with time the percentage of people adopting its use increased.

Some participants in this study don't want to change because they find the normal way of selling is much better and more profitable for them and they are not familiar with e-commerce. For example, Ali (1) commented "*we are familiar with the way of normal selling where a customer has to come to our shop and see the products*".

Similarly with Hassan (1) where he clearly indicated that "*e-commerce is not similar to normal way of selling where a customer comes to the shop, see a product, and make sure he/she selects the right product*".

All the cases mentioned that they are not familiar with e-commerce and have no sufficient knowledge to use e-commerce were in the e-commerce maturity levels 1 and 2. Within company, insufficient knowledge to deal with e-commerce was significant inhibitor for these companies to use e-commerce. Hassan (1) mentioned that "*our staffs don't know about marketing and selling on the Internet where they required training programs*". Similarly with Nasir (2) when he was asked what prevent his company not to use e-commerce, he replied "*honestly there is Ignorance inside our company in terms of e-commerce, and that is why our website only displays products and not receives orders*". To go ahead with e-commerce, they are required to train their employees and employ other skilled people. In some cases, this is seen costly, not useful and not fitted with their business priorities.

4.3.5. Business Priority

Business priority can be defined as "a statement of the level or urgency of important business needs" (Suppiah, not dated). Participating businesses which mentioned the business priority, they stated that e-commerce is not one of their current business priorities so they do not consider its implementation at the top their businesses. For example, Fadi's (2) company already has its website that show their products and prices, has telephone orders and own delivery system. It seems almost everything is ready for them to do e-commerce, the only thing that they need to activate receiving orders on their website. However, e-commerce is not considered in their business priorities at the moment. "*Using the Internet to order our products is not our priority at the moment*", Fadi (2) said.

Another example affecting business priority is consumers' purchasing power. A manufacturing company producing Islamic watches and clocks has e-commerce website mainly targets international market and has no priority to sell locally.

This is because *"the purchasing power comes from overseas and that's why this website is in English"*, Naif (2) said. They do not have priority to sell online locally because they considered that their products are plentiful in the Saudi market and there is no point to sell online locally. It is interesting case for discussion! While this business already has a channel selling online to international market, what prevent them to target local market on the same channel that they already they have? It seems that there are other factors other than business priority and products being available in the local market. Unfortunately, the researcher missed to emphasize on this interesting point to figure out what is really behind that decision.

A third cause for business priority not to do e-commerce is because the perceived low income that can be gained from selling online. That is the nature of business; business men are looking to maximize their profits. The idea or strategy that is seen bring more profits has higher priority and the opposite has less priority. Ahmed (2) replied on a question why not using e-commerce saying *"we have a different marketing strategy which is having branches/shops in the main malls to be directly close to customers"*. For Ahmed's (2) company e-commerce is seen not bring more profits whereas having physical shops and contacted with consumers in person do maximize their profits, it has higher priority.

4.3.6. Security and Trust Concerns

Concerns regarding security of using technologies have received considerable attention in the literature. These concerns become greater in an ineffective legal and regulatory environment (Dedrick, & Melville, 2006) and are the most significant barrier for technology non-adopters (Wymer & Regan, 2005). Similarly with this study, participating businesses which have concern regarding security and trust of e-commerce are non-adopters, levels 1 and 2 of e-commerce maturity model. From the statements made by interviewees, it appears that there is a lack of trust regarding online activities and that is linked to the lack of understanding within the business community. This lack of trust is primarily directed at online payment. For example, Ali (1) commented that *"we do not trust online payment; the money must paid cash"*. Participants who raised this issue had no experience selling and purchasing online, thus their concern is perceived. Some of these concerns are exaggerated. For example, Nasir (2) clearly stated that *"we don't trust online payment and our customers don't trust it either, the ignorance of something breeds fear from it"*. With the lack of knowledge the issues of trust and security increase.

Therefore, Nasir's perception about selling and buying over the Internet was not good because "*there are thefts of Credit Cards numbers, and there are hackers who penetrate your privacy*". On the Internet there are secure and trust payment methods that enormous business around the world using them. Again, the issue here is about the lack of knowledge and experience.

### 4.3.7. Setup and Maintenance Cost Concern

The cost of setting up and maintaining electronic business, including the website and IT skilled staff may make it difficult for businesses to adopt e-commerce. Analysis of the data from interviewees suggests that there is some perceived difficulty and cost associated with setting up an online retailing system.

For businesses in levels 1 and 2 it seems they have higher concern regarding the setup cost. For example, Saleh (1) stated that "*the website needs cost and even we go ahead to build an e-commerce website, it is not guaranteed to be successful*". That is right that they have major concern regarding the results of using e-commerce, because the cost involved is not only with regards to build a website. Using e-commerce for businesses in stage 1 and 2 of e-commerce maturity involves training programs, computer network, dedicated staff to take care of e-commerce activity, etc, whereas businesses in stages 3 and 4 of e-commerce maturity have no concerns because they well established in internal e-readiness.

Again, businesses here link the cost of spending and profits that will be return. For the businesses that have concerns regarding setup and maintenance they do not see that e-commerce will bring profits. For that reason they choose not to be involved financially "*in something that does not bring profit*", Waleed (1) said.

This is one of the aspects where taxation law is useful in many countries. If e-commerce is a priority, then that country could do two things to make adoption less of a burden: create tax concessions, cover the initial losses that may be incurred. It appears that perception of set-up costs may be based on ignorance. There are several online companies that aid in getting a business online by providing business templates that can be quickly configured to the business' needs, and get the business online. Setup costs are generally replaced with ongoing costs in this model by paying for an online business hosting service that incorporates all of the payment, advertising, etc.

For somewhere like the KSA, where there is no taxation, concessions can still be made to offset business losses during adoption. This is where government support is vital to the promotion and adoption of this technology.

## 5. Conclusion

We have seen that the interview data has led to the development of 22 suggested factors influencing retailers in Saudi Arabia on whether or not to adopt e-commerce. Interestingly, the analysis came up with findings that have not been identified in the literature. It looks as though part of the problem comes from the retailers' perceptions (or in some cases, prejudgment) about consumers in Saudi Arabia. The more interesting thing is that there seem to be differences between retailers in companies in different maturity stages in terms of having different attitudes regarding the issues of using e-commerce. Having said that the interview data suggests that retailers in companies in different stages of e-commerce maturity have different attitudes, it is necessary to carry out a, quantitative investigation using many more retailers to examine whether there are significant differences between retailers in companies at different levels of maturity.